\documentclass[dvips]{arxstspdf}
\usepackage{graphicx}
\usepackage{flushend}
\usepackage{stfloats}

\volume{27}
\issue{1}
\pubyear{2012}
\firstpage{51}
\lastpage{60}
\doi{10.1214/10-STS319}

\makeatletter
\newtheorem{lemma}{Lemma}[section]
\newtheorem{theorem}[lemma]{Theorem}

\makeatother

\begin{document}
\begin{frontmatter}
\vspace*{6pt}
\title{Shrinkage Confidence Procedures}
\runtitle{Shrinkage Confidence Procedures}

\begin{aug}
\author[a]{\fnms{George} \snm{Casella}\corref{}\ead[label=e1]{casella@stat.ufl.edu}}
\and
\author[b]{\fnms{J. T. Gene} \snm{Hwang}\ead[label=e2]{hwang@math.cornell.edu}}
\runauthor{G. Casella and J. T. Gene Hwang}

\address[a]{George Casella is Distinguished Professor, Department of
Statistics, University of Florida,
Gainesville, FL 32611, USA \printead{e1}.}
\address[b]{J. T. Gene Hwang is Professor,
Department of Mathematics, Cornell University, Ithaca, NY 14853, USA,
and Adjunct Professor, Department of Statistics, Cheng Kung University,
Tainan, Taiwan \printead{e2}.}

\end{aug}

% ABSTRACT
%
\begin{abstract}
The possibility of improving on the usual multivariate normal
confidence was first discussed in Stein (\citeyear{Stein62}). Using the
ideas of
shrinkage, through Bayesian and empirical Bayesian arguments,
domination results, both analytic and numerical, have been obtained.
Here we trace some of the developments in confidence set estimation.

\end{abstract}

% KEYWORDS
%

\end{frontmatter}
%
%s1 ###
\section{Introduction}
In estimating a multivariate normal mean, the usual $p$-dimensional
$1-\alpha$ confidence set is
%
%e1 ###
%
\begin{equation}\label{eq:usual}
C_{x,\sigma}^{0}=\{\theta\dvtx | \theta-x| \le c\sigma\},
\end{equation}
where we observe $X=x$, where $X$ is a random varia\-ble with a
$p$-variate normal distribution with mean~$\theta$ and covariance
matrix $\sigma^2I$, $X \sim N(\theta, \sigma^2I)$, $I$ is the $p
\times p$ identity matrix, and $c^2$ is the upper $\alpha$ cutoff of a
chi-squared distribution, satisfying $P(\chi^2_p \le c^2)=1-\alpha$.

%Points to note:\but{Edit points to note}
%on a linear model can be reduced to $C_{x,\sigma}^{0}$ with the usual
%unbiased estimator $s^2$ substituted for $\sigma^2$. This is the usual
%Scheff$\acute{\mathrm{e}}$ confidence set.
%%\item We will mostly consider the case of $\sigma^2$ known, where
%there have been theoretical results.
%model where we have $\hat\beta\sim N(\beta, \sigma^2 \Sigma)$, where
%$\beta$ is the regression parameter and $\Sigma$ is the covariance
%matrix (usually depending on the design matrix). The typical
%confidence set for $\beta$ is
%c^2 \sigma^2 \}.

Although the above formulation looks somewhat naive, it is very
relevant in applications of the linear model, still one of the most
widely-used statistical models. For such models, typical assumptions
lead to $\hat\beta\sim N(\beta, \sigma^2 \Sigma)$, where $\hat
\beta$
is the least squares estimator (and MLE under normality), $\beta$ is
the vector of regression slopes and $\Sigma$ is a known covariance
matrix (typically depending on the design matrix). The usual confidence
set for $\beta$ is
%
%e2 ###
%
\begin{equation}\label{eq:usualregression}
\{\beta\dvtx (\hat\beta-\beta)^\prime\Sigma^{-1}(\hat\beta-\beta)
\le
c^2 \sigma^2 \}.
\end{equation}
Letting $x=\Sigma^{-1/2} \hat\beta$ and $\theta= \Sigma
^{-1/2}\beta$
reduces (\ref{eq:usualregression})\break to~(\ref{eq:usual}).

In theoretical investigations of confidence sets and procedures, we
often first take $\sigma^2$ known. When $\sigma^2$ is unknown, the
usual strategy is to replace it by some usual estimator, such as the
sample variance~$s^2$. Under normality, if $s^2$ has $\nu$ degrees of
freedom, then $s^2 \sim\sigma^2 \chi^2_\nu$, independent of $\hat
\beta$. For example, the usual~$F$ confidence set for the regression
parameters based on a linear model can be reduced to $C_{x,\sigma}^{0}$
with the usual unbiased estimator $s^2$ substituted for $\sigma^2$.
This is the usual Scheff$\acute{\mathrm{e}}$ confidence set.
Unfortunately, contrary to the point estimation case, there are few
theoretical results for unknown $\sigma^2$. However, there is continued
numerical evidence that the usual confidence set can be dominated in
the unknown variance case (see, e.g., Casella and Hwang, \citeyear{Casella87}).
Moreover, Hwang and Ullah (\citeyear{Hwang94}) argue that the domination of the
alternative fixed radius confidence spheres for the unknown~$\sigma^2$
case, over Scheff\'{e}'s set, holds with a larger shrinkage factor.

Since we are assuming that $\sigma^2$ is known, we take it equal to $1$
and (\ref{eq:usual}) becomes
%
%e3 ###
%
\begin{equation}\label{eq:usual2}
C_{x}^{0}=\{\theta:| \theta-x| \le c\}.
\end{equation}

We now ask the question of whether it is possible to improve on $C_x^0$
in the sense of finding a confidence set $C^\prime$ such that, for all
$\theta$ and $x$:
\begin{longlist}[(ii)]
\item[(i)] $P_\theta(\theta\in C^\prime) \ge P_\theta(\theta\in
C_{x}^{0})$;
\item[(ii)] $\mbox{ volume of }C^\prime\le\mbox{ volume of }C_{x}^{0}$;
\end{longlist}
with strict inequality holding in either (i) or (ii) for a set
$\theta$ or $x$ with positive Lebesgue measure. The answer to this
question may be yes for higher dimensional cases, as suggested by the
work of Stein.

The celebrated work of James and Stein (\citeyear{James61}) shows that the estimator
%
%e4 ###
%
\begin{equation}\label{eq:JS}
\delta^{\mathrm{JS}}(x) = \biggl(1-\frac{a}{|x|^2}\biggr)x
\end{equation}
dominates $X$ with respect to squared error loss if $0 < a < 2(p-2)$,
that is,
%
%e5 ###
%
\begin{equation}\label{eq:JSdom}
\qquad\mathrm{E}_\theta| \delta^{\mathrm{JS}}(X)-\theta|^2 \cases{
\le\mathrm{E}_\theta|X-\theta|^2& $\mbox{for all }
\theta,$\vspace*{2pt}\cr
< \mathrm{E}_\theta|X-\theta|^2 & $\mbox{for some } \theta.$}
\end{equation}
In practice, this estimator has the deficiency of a~singularity at $0$
in that $\lim_{|x| \rightarrow0} \delta^{\mathrm{JS}}(x)=- \infty$. This
deficiency can be corrected with the positive part estimator (appearing
in Baranchik, \citeyear{Baranchik64}, and mentioned as Example 1 in Baranchik, \citeyear{Baranchik70})
%
%e6 ###
%
\begin{equation}\label{eq:JSpos}
\delta^{+}(x) = \biggl(1-\frac{a}{|x|^2}\biggr)^+ x,
\end{equation}
where $(b)^+=\max\{0,b\}$. This estimator actually improves on $\delta
^{\mathrm{JS}}(x)$ and is so good that, even though~it was known to be
inadmissible, it took $30$ years~to find a dominating estimator (Shao
and \mbox{Strawderman}, \citeyear{Shao94}). The removal of the singularity makes~$\delta
^{+}(x)$ a~more attractive candidate for centering a~confidence set.

A simple proof of (\ref{eq:JSdom}) can be found in Stein (\citeyear{Stein81}); see
also Lehmann and Casella (\citeyear{Lehmann98}), Chapter 5. Therefore, it seems
reasonable to conjecture that we can use a Stein estimator to dominate
the confidence set $C_x^0$. Although this turns out to be the case, it
is a very difficult problem.

%s2 ###
\section{Recentering}

Stein (\citeyear{Stein62}) gave heuristic arguments\footnote{Stein's paper must be
read carefully to appreciate these arguments. He uses a large $p$
argument and the fact that $X$ and $X-\theta$ are orthogonal as $p
\rightarrow\infty$.} that sho\-wed why
recentered sets of the form
%
%e7 ###
%
\begin{equation}\label{eq:pospartset}
C_\delta^+ = \{\theta\dvtx | \theta- \delta^+
(\mathbf{x})
| \leq c\}
\end{equation}
would dominate the usual confidence set (\ref{eq:usual2}) in the sense
that $ P_\theta(\theta\in C_\delta^+ (\mathbf{X})) >
P_\theta
(\theta\in C_x^0
(\mathbf{X})) $ for all~$ \theta$, where $ \mathbf{X} \sim N
(\theta, I) , p \geq3 $. (Note that this set has the same
volume as
$C_x^0$, but is recentered $\delta^+$. Dominance would thus be
established if we can show that $C_\delta^+$ has higher coverage
probability than $C_x^0$.) Stein's argument was
heuristic, but Brown (\citeyear{Brown66}) and Joshi (\citeyear{Joshi67}) proved the inadmissibility
of $
C_x^0 $ if $ p \geq3 $ (without giving an explicit dominating
procedure). Joshi
(\citeyear{Joshi69}) also showed that $ C_x^0 $ was admissible if $ p \leq2 $.\vadjust{\goodbreak}

The existence results of Brown and Joshi are based on spheres centered at
%
%e8 ###
%
\begin{equation}\label{eq:BrownJoshiEstimator}
\biggl( 1-\frac{a}{b+|x|^2}\biggr) x
\end{equation}
[compare to (\ref{eq:JSpos})] where $a$ is made arbitrarily small and
$b$ is made arbitrarily large. But these existence results fall short
of actually exhibiting a confidence set that dominates $C_x^0$.

The first analytical and constructive results were established by
(surprise!) Hwang and Casella (\citeyear{Hwang82}), who studied the coverage
probability of $C_\delta^+$ in (\ref{eq:pospartset}). Since $C_\delta
^+$ and $C_x^0$ have the same volume, domination will be established if
it can be shown that $C_\delta^+$ has higher coverage probability for
every value of $\theta$. It is easy to establish that:
\begin{itemize}
\item[$\circ$] $P_\theta(\theta\in C_\delta^+(X))$ is only a function
of $|\theta|$, the Euclidean norm of $\theta$, and
\item[$\circ$] $\lim_{|\theta| \rightarrow\infty} P_\theta
(\theta\in
C_\delta^+(X))=1-\alpha$, the coverage pro\-bability of $C_x^0$.
\end{itemize}
Therefore, to prove the dominance of $C_\delta^+$, it is sufficient to
show that the coverage probability is a nonincreasing function of
$|\theta|$. Hwang and Casella (\citeyear{Hwang82}) derived a formula for
$(d/d|\theta
|)P_\theta(\theta\in C_\delta^+(X))$ and found a constant $a_0$
(independent of $\theta$) such that if $0<a<a_0$, $C_\delta^+$
dominates $C_x^0$ in coverage probability for $p \ge4$. Using a
slightly different method of proof, Hwang and Casella (\citeyear{Hwang84}) extended
the dominance to cover the case $p=3$. This proof is outlined in
Appendix \hyperref[app:hc]{A}. The analytic proof was generalized to spherical
symmetric
distributions by Hwang and Chen (\citeyear{Hwang86}).

There is an interesting geometrical oddity associated with the Stein
recentered confidence set. To see this, we first formalize our
definitions of confidence sets. Note that for any confidence set we can
speak of the $x$-section and the $\theta$-section. That is, if we
define a \textit{confidence procedure} to be a set $C(\theta,x)$ in the
product space $\Theta\times\mathcal{X}$, then:
\begin{enumerate}[(1)]
\item[(1)] The $x$-section, $C_x=\{\theta\dvtx\theta\in C(\theta,x)\}$, is the
confidence set.
\item[(2)] The $\theta$-section, $C_\theta=\{x\dvtx x \in C(\theta,x)\}$, the
acceptance region for the test $H_0\dvtx\{\theta\}$.
\end{enumerate}
We then have the tautology that $\theta\in C _x$ if and only if $x \in
C_\theta$ and, thus, we can evaluate the coverage probability
$P_\theta
(\theta\in C_X)$ by computing $P_\theta(X \in C_\theta)$, which is
often a more straightforward calculation.

\begin{figure}

\includegraphics{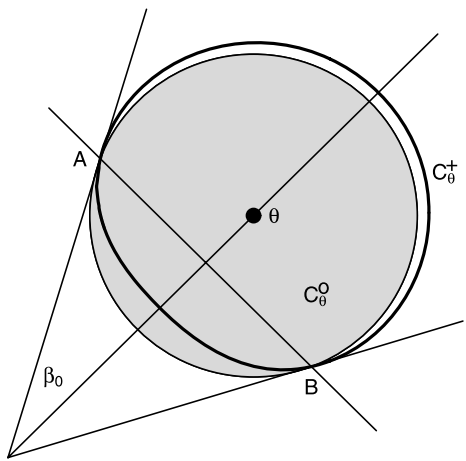}

\caption{Two-dimensional representation for
$C_\theta
^+$ and $C_\theta^0$ for $|\theta|>c$, where $C_\theta^0$ is the sphere
of radius $c$ centered at $\theta$ (shaded). The set $C_\theta^+$
intersects $C_\theta^0$ at point $A$ and $B$ (details on the points of
intersection are in Hwang and Casella, \protect\citeyear{Hwang82}). Note the flattening of
$C_\theta^+$ on the side toward the origin and the decrease in volume
over $C_\theta^0$.}\label{fig:JSSet}
\end{figure}

For the usual confidence set, both $C_x^0$ and $C_\theta^0$ are
spheres, one centered at $x$ and one centered at $\theta$. Although the
confidence set $C_x^+$ is a sphere, the associated $\theta$-section\vadjust{\goodbreak}
$C_\theta^+$ is not, and has the shape portrayed in Figure \ref
{fig:JSSet}. Notice the flattening of the set in the side closer to $0$
in the direction perpendicular to~$\theta$, and the slight expansion
away from $0$. Stein (\citeyear{Stein62}) knew of this flattening phenomenon, which
he noted can be achieved in any fixed direction. What is interesting is
that this reshaping of the $\theta$-section of the recentered set leads
to a set with higher coverage probability than $C_x^0$ when $p \ge3$.

%s3 ###
\section{Recentering and Shrinking the~Volume}\label{sec:recentershrink}

The improved confidence sets that we have discussed thus far have the
property that their coverage
probability is uniformly greater than that of~$ C_x^0 $, but the
infimum of the
coverage probability (the \textit{confidence coefficient}) is equal to
that of $ C_x^0 $. For example, recentered sets such as $ C_\delta^+$
will present the same volume and confidence coefficient to an
experimenter so, in practice, the experimenter has not gained anything.
(This is, of course, a fallacy and a~shortcoming of the frequentist
inference, which requires the reporting of the infimum of the coverage
probability.)

However, since the coverage probability of $C_\delta^+$ is uniformly
higher than the infimum $\inf_\theta P_\theta(\theta\in
C_X^0)=1-\alpha
$, it should be possible to reduce the radius of the recentered set and
maintain dominance in coverage probability.

%This is the approach of Casella and Hwang (1983, 1987) who...{\bf need
%to finish}

%{\bf What other papers should go here??}

In this section we describe some approaches to constructing improved
confidence sets, approaches that not only result in a recentering of
the usual set, but also try to reduce the radius (or, more generally,
the volume). Some of these constructions are based on variations of
Bayesian highest posterior density regions, and thus share the problem
of trying to describe exactly what the $x$-section, the confidence set,
looks like. Others are more of an empirical Bayes approach, and tend to
have more transparent geometry.

%s3.1 ###
\subsection{Reducing the Volume--Bayesian Approaches}

The first attempt at constructing confidence sets with reduced volume
considered sets with the same coverage probability as $C_X^0$, but with
uniformly smal\-ler volume. One of the first attempts was that of Faith
(\citeyear{Faith76}), who considered a Bayesian construction based on a two-stage
prior where
\[
\theta\sim N(0, t^2I),\quad  t^2 \sim\mbox{Inverted }\operatorname{Gamma}(a,b),
\]
which is similar (but not equal) to the prior used by Strawderman
(\citeyear{Strawderman71}) in the point estimation problem (Appendix \hyperref[app:straw]{B}). The
two-stage prior amounts to a proper prior with density
\[
\pi(\theta) \propto( 2b+|\theta|^2)^{-(a+p/2)},
\]
the multivariate $t$-distribution with $2a$ degrees of~free\-dom. Faith
then derived the Bayes decision against a linear loss, but modified it
to the more explicitly defined region
\[
C_F = \biggl\{\theta\dvtx\biggl(\frac{\exp(c^2)}{\exp(|x-\theta|^2)}
\biggr)^{1/(p+2a)} \ge\frac{2b+\theta^2}{2b+|x|^2}\biggr\},
\]
where $c$ is the radius of $C_x^0$. It may happen that $C_F$ is not
convex. However, if $a>-p/2$ and $b>(a+p/2)/8$, the convexity of $C_F$
was established. Unfortunately, little else was established except when
$p=3$ or $p=5$, where for some ranges of $a$ and $b$ it was shown that
$C_F$ has smaller volume and higher coverage probability than $C_x^0$.

Berger (\citeyear{Berger80}) took a different approach. Using a~generalization of
Strawderman's prior, he calculated the posterior mean $\delta_B(x)$ and
posterior covariance matrix $\Sigma_B(x)$ and recommended
\[
C_B = \bigl\{\theta\dvtx\bigl(\theta-\delta_B(x)\bigr)^{\prime}\Sigma
_B(x)^{-1}\bigl(\theta-\delta_B(x)\bigr) \le\chi^2_{p,\alpha}\bigr\},
\]
where $\chi^2_{p,\alpha}$ is the upper $\alpha$ cutoff point from a
chi-square distribution with $p$ degrees of freedom. The posterior
coverage probability would be exactly \mbox{$1-\alpha$} if the posterior
distribution were normal, but this is not the case (and the posterior
coverage is not the frequentist coverage). However, Berger was able to
show that his set has very attractive coverage probability and small
expected volume based on partly analytical and partly numerical evidence.

%s3.2 ###
\subsection{Reducing the Volume--Empirical Bayes Approaches}

A popular construction procedure for finding good point estimators is
the empirical Bayes approach (see Lehmann and Casella, \citeyear{Lehmann98}, Section
4.6, for an introduction), and proves to also be a useful tool in
\mbox{confidence} set construction. However, unlike the point estimation
problem, where a direct application of empirical Bayes arguments led to
improved Stein-type estimators (see, e.g., Efron and Morris,
\citeyear{Efron73}), in the confidence set problem we find that a straightforward
implementation of an empirical Bayes argument would not result in a
$1-\alpha$ confidence set. Modifications are necessary to achieve
dominance of the usual confidence set.

Suppose that we begin with a traditional normal prior at the first
stage, and have the model
\[
X \sim N(\theta, I),\quad  \theta\sim N(0, \tau^2I),
\]
which results in the Bayesian Highest Posterior Density (HPD) region
%
%e9 ###
%
\begin{equation}\label{eq:hpd}
C^\pi=\{\theta\dvtx |\theta- \delta^\pi(x)|^2 \le c^2 M
\},
\end{equation}
where $M=\tau^2/(\tau^2+1)$ and $\delta^\pi(x)=Mx$ is the Bayes point
estimator of $\theta$. This follows from the classical Bayesian result
that $\theta\vert x \sim N(Mx,MI)$.

However, for a fixed value of $\tau$, the set $C^\pi$ cannot have
frequentist coverage probability above $1-\alpha$ for all values of
$\theta$. This is easily seen, as the posterior coverage is identically
$1-\alpha$ for all $x$, and, hence, the double integral over $x$ and
$\theta$ is equal to $1-\alpha$. This means that the frequentist
coverage is either equal to $1-\alpha$ for all $\theta$, or goes above
and below $1-\alpha$. Since the former case does not hold (check
$\theta
=0$ and a nonzero value), the coverage probability of $C^\pi$ is not
always above $1-\alpha$.

Consequently, if we take a naive approach and replace $\tau^2$ by a
reasonable estimate, an empirical Bayes approach, we cannot expect that
such a set would maintain frequentist coverage above $1-\alpha$. This
is because such a set would have coverage probabilities converging to
those of $C^\pi$ (as the sample size increases) and, hence, such an
empirical Bayes set would inherit the poor coverage probability of~$C^\pi$. This phenomenon has been documented in Casella and Hwang (\citeyear{Casella83}).

As an alternative to the naive empirical Bayes approach, consider a
decision-theoretic approach with a loss function to measure the loss of
estimating the parameter $\theta$ with the set $C$:
%
%e10 ###
%
\begin{equation}\label{eq:loss}
L(\theta, C) = k\operatorname{vol}(C) - I(\theta\in C),
\end{equation}
where $k$ is a constant, $\operatorname{vol}(C)$ is the volume of the set~$C$,
and $I(\cdot)$ is the indicator function. Starting with a prior
distribution $\pi(\theta)$, the Bayes rule against $L(\theta, C)$ is
the set
%
%e11 ###
%
\begin{equation}\label{eq:hdregion}
\{ \theta\dvtx \pi(\theta\vert x) > k\},
\end{equation}
where $\pi(\theta\vert x)$ is the posterior distribution. This is a~highest posterior density (HPD) region.

The choice of $k$ is somewhat critical, and we chose it to coincide
with properties of $C^0$. Specifically, if we chose $k=\exp
(-c^2/2)/(2\pi)^{p/2}$, then $C^0$ is minimax for the loss (\ref
{eq:loss}). An alternative explanation of this choice of $k$ is based
on the reasoning that as $\tau\rightarrow\infty$, (\ref{eq:hdregion})
would converge to $C^0$, which insures that the alternative intervals
would not become inferior to $C^0 $ for large $\tau^2$. (See He, \citeyear{He92};
Qiu and Hwang, \citeyear{Qiu07}; and Hwang, Qiu and Zhao, \citeyear{Hwang08}.) Applying this
choice of $k$ with the normal prior $\theta\sim N(0, \tau^2I)$
yields the Bayes set
\[
C^\pi_{x,k} = \{ \theta\dvtx \vert\theta- \delta^\pi(x) \vert\le
M[c^2-p \log M]\},
\]
where $ \delta^\pi(x) $ and $M$ are as in (\ref{eq:hpd}). By estimating
the hyperparameters, this is then converted to an empirical Bayes set
\[
C^E_{x} = \{ \theta\dvtx \vert\theta- \delta^{+}(x) \vert\le
v_E(x)\},
\]
where $\delta^{+}(x)$ is the positive part estimator of (\ref
{eq:JSpos}), and $v_E(x)$ is given by
%
%e12 ###
%
\begin{eqnarray}\label{eq:ppv}
v_E(x) &=& \biggl(1-\frac{p-2}{\max(|x|^2, c^2)}\biggr)
\nonumber
\\[-8pt]
\\[-8pt]
\nonumber
&&{}\cdot\biggl[c^2-p\log
\biggl(1-\frac{p-2}{\max(|x|^2, c^2)}\biggr)\biggr].
\end{eqnarray}
When $c^2>p$, a minor condition requiring $1-\alpha > 0.55$,
$M[c^2-p\log M] \uparrow c^2$ as $M \rightarrow \infty$.
It also follows that $v_E(x)$ is bounded away from zero. This is important in maintaining
coverage probability. Extensive numerical evidence was given (Casella and Hwang, \citeyear{Casella83}) to support
the claim that $C^E_{x}$ is a~uniform improvement over $C^0_{x}$.

Confidence sets with exact $1-\alpha$ coverage probability, with
uniformly smaller volume, have also been constructed by Tseng and Brown
(\citeyear{Tseng97}), adapting results from Brown et al. (\citeyear{Brown95}). These
confidence sets are shown, numerically, to typically have smaller
volume that those of Berger (\citeyear{Berger80}).

Brown et al. (\citeyear{Brown95}), working on the problem of bioequivalence,
start with the inversion of an $\alpha$-level test and derive a
$1-\alpha$ confidence interval that minimizes a Bayes expected volume,
that is, the volume averaged with respect to both $x$ and $\theta$.
Tseng and Brown (\citeyear{Tseng97}), using a normal prior $\theta\sim N(0,
\tau^2I)$, show that the corresponding set of Brown et al.
(\citeyear{Brown95}) becomes
\[
C^\mathrm{B} = \biggl\{ \theta\dvtx \biggl\vert x-\theta\biggl( \frac{1+\tau^2}{\tau
^2} \biggr)\biggr\vert^2 \le k (\vert\theta\vert^2/\tau^4)
\biggr\},
\]
where $k(\cdot)$ is chosen so that $C^\mathrm{B}$ has exactly $1-\alpha$
coverage probability for every $\theta$. A simple calculation shows
that the squared term in $C^\mathrm{B}$ has a noncentral chi squared
distribution, so $k(\cdot)$ is the appropriate $\alpha$ cutoff point.
In doing this, Tseng and Brown avoided the problem of Casella and Hwang
(\citeyear{Casella83}), and the radius does not need to be truncated.

Of course, to be usable, we must estimate $\tau^2$. The typical
empirical Bayes approach would be to replace $\tau^2$ with an estimate,
a function of $x$. However, Tseng and Brown take a different approach
and replace $\tau^2$ with a function of $\theta$, thereby maintaining
the $1-\alpha$ coverage probability. They argue that $\theta$ is more
directly related to $\tau$ than is $x$, and should provide a better
``estimator.'' Examples of this approach are discussed in Hwang (\citeyear{Hwang95}) and Huwang (\citeyear{Huwang96}).

The set proposed by Tseng and Brown is
\begin{eqnarray*}
C^\mathrm{TB} &=& \biggl\{ \theta\dvtx \biggl\vert x-\theta\biggl( 1+\frac{1}{A + B
|\theta|^2} \biggr)\biggr\vert^2\\
&&\qquad{} \le k \biggl(\biggl(\frac{ |\theta|}{A + B
|\theta|^2}\biggr)^2\biggr)\biggr\}
\end{eqnarray*}
for constants $A \ge0$ and $B > 0$, and has coverage exactly equal to
$1-\alpha$ for every $\theta$. Combining analytical results and
numerical calculations, these sets are shown to have uniformly smaller
volume that $C_x^0$. Moreover, Tseng and Brown also demonstrate volume
reductions over the sets of Berger (\citeyear{Berger80}) and Casella and Hwang (\citeyear{Casella83}).
The only quibble with their approach is that the exact form of the set
is not explicit, and can only be solved numerically.

%s3.3 ###
\subsection{Reducing Volume and Increasing Coverage}

The first confidence set analytically proven to have smaller volume and
higher coverage than $C_x^0$ is that of Shinozaki (\citeyear{Shinozaki89}). Shinozaki
worked with the $x$-section of the confidence set, starting with the set~$C_x^0$. Consider Figure \ref{fig:JSSet}, but drawn as the $x$-section
centered at $x$. By shrinking $C_x^o$ toward the origin, he was able to construct a new set with the same coverage
probability as $C_x^0$ but smaller volume. These sets can have a
substantial improvement over $C_x^0$, but smaller improvements compared
to Berger (\citeyear{Berger80}) and Casella and Hwang (\citeyear{Casella83}) (especially when $p$ is
large and $\vert\theta\vert$ is small). Moreover, there is no point
estimator that is explicitly associated with this set.

%The confidence sets described in Section \ref{sec:recentershrink},
%which attain some volume reduction but maintain the same confidence
%coefficient as $ C^0 $ still are somewhat ``wasteful'' because they
%have coverage probabilities higher than that of $ C^0 $. However, this
%deficiency was overcome by Tseng and Brown (1997).

%By adapting results of Brown et al. (1995), Tseng and Brown
%(1997) constructed an improved confidence set, $C^\ast$, with the
%property that $ P_\theta(\btheta\in C^\ast({\bf X})) = P_\theta(
%amount of volume reduction while
%maintaining the same coverage probability as $ C^0 $.

%s3.4 ###
\subsection{Other Constructions}
Samworth (\citeyear{Samworth05}) looked at confidence sets of the form
\[
\{\theta\dvtx \vert\theta- \delta^{+} \vert^2 \le w_\alpha(\theta)
\},
\]
where $\delta^{+}$ is the positive part estimator (\ref{eq:JSpos}),
$w_\alpha(\theta)$ is the appropriate $\alpha$-level cutoff to give the
confidence set coverage probability $1-\alpha$ for all $\theta$, and~$X$ has a~spherically symmetric distribution. He then replaced~$w_\alpha(\theta)$ by its Taylor expansion
\[
w_\alpha(\theta) \approx w_\alpha(0) + \tfrac{1}{2}w_\alpha^{\prime
\prime}(0)|\theta|^2,
\]
and, replacing $\theta$ with $x$, arrived at the confidence set
\[
\bigl\{\theta\dvtx \vert\theta- \delta^{+} \vert^2 \le\min\bigl(w_\alpha
(0) + \tfrac{1}{2}w_\alpha^{\prime\prime}(0)|x|^2, c^2\bigr)\bigr\}.
\]
Samworth noted the importance of the quantity\break $f^\prime(c^2)/ f(c^2)$,
where $f$ is the density of $x$ (the relative increasing rate of $f$ at
$c^2$). The radius of the analytic confidence set only depends on the
density through $c^2$ and $f^\prime(c^2)/f(c^2)$. This point was
previously noted by Hwang and Chen (\citeyear{Hwang86}) and Robert and Casella (\citeyear{Robert90}).

This confidence set compares favorably with that of Casella and Hwang
(\citeyear{Casella83}), having smaller volume especially when $|x|$ is small. Numerical
results were given not only for the normal distribution, but also for
other spherically symmetric distributions such as the multivariate $t$
and the double exponential. Furthermore, a parametric bootstrap
confidence set is also proposed, which also performs well.

Efron (\citeyear{Efron06}) studies the problem of confidence set construction with
the goal of minimizing volume. He ultimately shows that seeking to
minimize volume may not be the best way to improve inferences, and that
relocating the set is more important than shrinking it. Using a unique
construction based on a polar decomposition of the normal density,
Efron derived a ``confidence density'' which he used to construct sets
with $1-\alpha$ coverage probability, and ultimately a minimum volume
confidence set with $1-\alpha$ posterior probability.

The confidence density, which plays a large part in Efron's paper, is
used to show the importance of locating the confidence set properly.
The sets of Tseng and Brown (\citeyear{Tseng97}) and Casella and Hwang (\citeyear{Casella83}) perform
well on this evaluation. A minimum volume construction is also derived,
and it is shown that the resulting set is not optimal in any
inferential sense. Inferential properties, similar to type I and type
II errors, are explored. It is also seen that as the relocated sets
decrease volume of the confidence set, they \textit{increase} the
acceptance regions.

%s4 ###
\section{Shrinking the Variance}\label{sec:varshrink}

Thus far, we have only addressed the problem of improving confidence
regions for the mean. However, there is also a Stein effect for the
estimation of the variance, and this can be exploited to produce
improved confidence intervals for the variance.

Stein (\citeyear{Stein64}) was the first to notice this (of course!). Specifically,
let $X_1, \ldots, X_n$ be i.i.d. $N(\mu, \sigma^2)$, univa\-riate, where
both $\mu$ and $\sigma$ are unknown, and calcula\-te $\bar X=(1/n) \sum_i
X_i$ and $S^2=\sum_i(X_i-\bar X)^2$. Against squared error loss, the
best estimator of $\sigma^2$, of the form $c S^2$, has $c=(n+1)^{-1}$.
This is also the best equivariant estimator [with the location-scale
group and the equivariant loss $(\delta-\sigma^2)^2/\sigma^4$], and is
minimax. Stein showed that the estimator
\begin{eqnarray*}
\delta^S(\bar X, S^2)&=& h( \bar X^2/S^2)S^2,\\
 h( \bar X^2/S^2) &=&\min
\biggl\{ \frac{1}{n+1}, \frac{1+n\bar X^2/S^2}{n+2}\biggr\} ,
\end{eqnarray*}
uniformly dominates $ S^2/(n+1)$. Notice that $\delta^S(\bar X,\break S^2)$
converges to $ S^2/(n+1)$ if $\bar X^2/S^2$ is big, but shrinks the
estimator toward zero if it is small. Stein's proof was quite
innovative (and is reproduced
in the review paper by Maatta and Casella, \citeyear{Maatta90}). The proof is based on
looking at the conditional expectation of the risk function,
conditioning on $\bar X/S$, and showing that moving the usual estimator
toward zero moves to a lower point on the quadratic risk surface. This
approach was extended by Brown (\citeyear{Brown68}) to establish inadmissibility
results, and by Brewster and Zidek (\citeyear{Brewster74}), who found the best scale
equivariant estimator.
Minimax estimators were also found by Strawderman (\citeyear{Strawderman74}), using a
different technique.

Turning to intervals, building on the techniques developed by Stein
and\vadjust{\goodbreak}
Brown, Cohen (\citeyear{Cohen72}) exhibited a confidence interval for the variance
that improved on the usual confidence interval. If $(S^{2}/b,\break S^{2}/a)$
is the shortest $1-\alpha$ confidence interval based on $S^{2}$ (Tate
and Klett, \citeyear{Tate59}), Cohen (\citeyear{Cohen72}) considered the confidence interval
\begin{eqnarray*}
&&(S^{2}/b, S^{2}/a)I( \bar X^2/S^2 > k)\\
&&\quad{} + (S^{2}/b^\prime,
S^{2}/a^\prime)I( \bar X^2/S^2 \le k ),
\end{eqnarray*}
where $I(\cdot)$ is the indicator function, $1/a-1/b = 1/a^{\prime
}-1/b^{\prime}$, so each piece has the same length, but $1/a^{\prime
}<1/a$ and $1/b^{\prime}<1/b$. So if $\bar X^2/S^2$ is small, the
interval is pulled toward zero, analogous to the behavior of the Stein
point estimator. Shorrack (\citeyear{Shorrock90}) built on this argument, and those of
Brewster and~Zi\-dek (\citeyear{Brewster74}), to construct a generalized Bayes confiden\-ce
interval that smoothly shifts toward zero, keeping the same length as
the usual interval but uniformly increasing coverage probability.
Building further on these arguments, Goutis and Casella (\citeyear{Goutis91})
constructed generalized Bayes intervals that smoothly shifted the usual
interval toward zero, reducing its length but maintaining the same
coverage probability. For more recent developments on variance
estimation see Kubokawa and Srivastava (\citeyear{Kubokawa03}) and Maruyama and
Strawderman (\citeyear{Maruyama06}).

%s5 ###
\section{Confidence Intervals}

In some applications there may be interest in making inference
individually for each $\theta_i$. One example is the analysis of
microarray data in which the interest is to determine which
genes are differentially expressed (i.e., having $\theta_i$, the
difference of the true expression between the
treatment group and the control group, different from zero). Although
the confidence sets of the previous section can be projected to obtain
confidence intervals, that will typically lead to wider intervals than
a direct construction.

If $X_i$ are i.i.d. $N(\theta_i,\sigma_i^2)$,
$i=1,\ldots,p$, the usual one-dimensional interval is
\[
I_{X_i}^0=X_i\pm c\sigma_i,
\]
where $c$ is chosen so that the coverage probability is $1-\alpha$.
Hence, $c$ is the $\alpha/2$
upper quantile of a~standard normal.

%s5.1 ###
\subsection{Empirical Bayes Intervals}

If a frequentist criterion is used, it is not possible to
simultaneously improve on the length and coverage probability of
$I_{X_i}^0$ in one dimension. However, it is
possible to do so if an empirical Bayes criterion is used.
Morris (\citeyear{Morris83}) defined an empirical Bayes confidence region with respect
to a class of priors $\Pi$, having confidence coefficient $1-\alpha$
to be a set $C(X)$ satisfying
\begin{eqnarray*}
P_\pi\bigl(\theta\in C(X)\bigr)&=&\int P_\theta\bigl(\theta\in
C(X)\bigr)\pi(\theta)\,d\theta\\
&\ge&1-\alpha\quad \mbox{for all } \pi(\theta)
\in
\Pi.
\end{eqnarray*}

Note that $P_\pi(\theta\in C(X))$ is the Bayes coverage
probability in that both $X$ and $\theta$ are integrated out. Using
normal priors with both equal and unequal
variance, Morris went on to construct $1-\alpha$ empirical Bayes confidence
intervals that have average (across $i$) squared lengths smaller than
$I_X^0$. Bootstrap intervals based on Morris' construction are also
proposed in Laird and Louis (\citeyear{Laird83}).

In the canonical model
%
%e13 ###
%
\begin{eqnarray}\label{model:He}\label{sec:interval:He}
X_i &\sim& \mathrm{i.i.d.}\ N(\theta_i,1) \quad \mbox{and}
\nonumber
\\[-8pt]
\\[-8pt]
\nonumber
\theta_i &\sim&
\mathrm{i.i.d.}\
{N}(0,\tau^2),
\end{eqnarray}
He (\citeyear{He92}) proved that there exists an
interval that dominates $I_X^0$. Precisely, for $\delta^{+}(X)$ of
(\ref
{eq:JSpos}), it was shown that there
exists $a>0$ such that the interval $\delta_i^+(X)\pm c$ has higher
Bayes coverage probability for any
$\tau^2>0$.

The approach He took is similar to the approach of Casella and Hwang
(\citeyear{Casella83}), using a one-dimensional loss function similar to the
linear loss (\ref{eq:loss}) except that $\theta$ is replaced by only
the component $\theta_i$ of interest. As in the discussion following
(\ref{eq:loss}),
$k$ and $c$ need to be properly linked. With such a choice of $k$, the
decision Bayes interval
is then approximated by its empirical Bayes counterpart:
\[
C_X^{\mathrm{He}}=\{\theta_i\dvtx |\theta_i-\delta_i^+(X)|^2\le\nu(|X|)\}.
\]
Here $\delta_i^+(X)$ is the $i$th component of the James--Stein
positive part estimator
(\ref{eq:JSpos}) with $a=p-2$,
%
%e14 ###
%
\begin{eqnarray}\label{eq:hwangint}
\nu(|X|)&=&\hat M(c^2-\log{\hat M}),
\nonumber
\\[-8pt]
\\[-8pt]
\nonumber
\hat M&=&\max\biggl\{ \biggl(1-\frac{p-2}{|X|^2}\biggr)^+,\frac
{1}{p-1}\biggr\}.
\end{eqnarray}
Note the resemblance to (\ref{eq:ppv}). There is also a
truncation carried out in the definition of $\hat M$ so that
$\nu(|X|)$ is bounded away from zero.

It can be shown that the length of $C_X^{\mathrm{He}}$ is
always smaller than that of $I_X^0$ for each individual coordinate,
$i$~as long as $c>1$, or, equivalently, $1-\alpha>68\%$. In contrast, in
Morris (\citeyear{Morris83}) only the average length across $i$ was made smaller.

%He's interval is compared with that of Morris (1983) and Laird and
%Louis (1987). For the canonical model (\ref{model:He}), Morris'
%interval ca be stated easily as
%I_X^M=\{\theta:|\theta-(1-\hat{B})X_i|^2\le[(1-\hat{B})+\frac{2X_i^2
%where $\hat{B}=\min(\frac{p-2}{p},\frac{p-2}{|X|^2})$.

%Note that the term $(1-\hat{B})$ on the right hand side of
%(\ref{interval:Morris}) equals $\max(2/p,\hat{M})$ where
%$\hat{M}=(1-\frac{p-2}{|X|^2})$. Hence Morris truncated $\hat{M}$ by
%$2/p$.

%Not only that Morris also add the term involving $X_i^2$ on the
%right hand side of (\ref{interval:Morris}). All these would help to
%improve on the low coverage probability, perhaps at the cost of
%being too wide. As the result of the second term, the interval
%length could be much shorter than $I_{X_i}^0$ where $X_i$ is large.
%(Interestingly, Morris also truncates the $\hat{B}$ in the point
%estimator. This would be connected in the next subjection.)

%He's interval adds only a logarithmic term in $\nu(|X|)$ which
%never larger than that of $I_{X_i}^0$.

Numerical studies in He (\citeyear{He92}) demonstrated that his interval is an
empirical Bayes confidence interval with $1-\alpha$ confidence
coefficient. Also, on average, it has shorter length than the
intervals of Morris (\citeyear{Morris83}) or Laird and Louis (\citeyear{Laird83}) when $\alpha
=0.05$ or
$0.1$. He concluded that his interval is recommended only if
$\alpha\le0.1$. Interestingly, in modern application with the
concerns of multiple testings, a small value of $\alpha$ is more important.

%s5.2 ###
\subsection{Intervals for the Selected Mean}

An important problem in statistics is to address the confidence
estimation problem after selecting a~subset of populations from a
larger set. This is especially so if the
number $p$ of populations is huge and the number of selected
populations, $k$, is relatively small, a scenario typical in microarray
experiments. For example, ignoring the selection and
just estimating the parameters of the selected populations by the
sample means would have serious bias, especially
if the populations selected are the ones with largest corresponding sample means. In
such a~situation, intuition would suggest that some kind of
shrinkage approach is very much needed.

Specifically, we consider the canonical model
%
%e15 ###
%
\begin{eqnarray}\label{model:QiuHwang}
X_i &\sim& \mbox{i.i.d. } N(\theta_i,\sigma_i^2) \quad \mbox{and}
\nonumber
\\[-8pt]
\\[-8pt]
\nonumber
\theta_i
&\sim&
\mbox{i.i.d. }N(\mu,\tau^2).
\end{eqnarray}
Let $\theta_{(i)}$ be the parameter of the selected population, that is,
it is the $\theta_j$ such that $X_j=X_{(i)}$ where
%
%e16 ###
%
\begin{equation}\label{inequality:QiuHwang}
X_{(1)}\le X_{(2)}\le\cdots\le X_{(p)}
\end{equation}
are the order statistics of $(X_1, \ldots, X_p)$. In particular,
$\theta
_{(p)}$ is
the $\theta$ that corresponds to the largest observation
$X_{(p)}=\max_jX_j$. Note that it is \textit{not true} that
$\theta_{(1)}\le\theta_{(2)}\le\cdots\le\theta_{(p)}$. In
particular, $\theta_{(p)}$ is not necessarily the largest of the
$\theta_j$'s. It is just that $\theta_j$ happens to have produced
the largest observations among the $X_i$'s.

%For further clarification, see Qiu and Hwang (2007).

In the point estimation problem, the naive estimator of $\theta_{(p)}$
is $X_{(p)}$, which can be
intuitively seen to be an overestimate, especially if all
$\theta_i$ are equal. A~shrinkage estimator adapted to this situation
would seem
more reasonable. Hwang (\citeyear{Hwang93}) was able to show that for estimating
$\theta_{(p)}$, a variation of the posi\-tive-part estimator (\ref
{eq:JSpos}), with $X_i$ replaced by $X_{(i)}$, has, for every $\mu$ and $\tau^2$, smaller Bayes risk
than $X_{(i)}$ with respect to one-dimensional squared error loss.

For the construction of confidence intervals, Qiu and Hwang (\citeyear{Qiu07})
adapted the approach of
Casella and Hwang (\citeyear{Casella83}) and He (\citeyear{He92}) to this problem.
For any selection, they constructed $1-\alpha$ empirical Bayes
confidence intervals for
$\theta_{(i)}$ which are shown numerically to have confidence
coefficient $1-\alpha$ when $\sigma_i=\sigma$ is either known
or estimable. Moreover, the interval is everywhere shorter than even
the traditional interval, $X_{(i)}\pm c\sigma$, which does not maintain
$1-\alpha$ coverage in this case.
%One difficulty is the truncation of the radius, restricting the length
%to
%above a positive number (as in (\ref{eq:ppv}) and (
%not work for this problem, and an alternate approach was developed.

Interestingly, in one microarray data set, Qiu and Hwang (\citeyear{Qiu07}) found
that the
normal prior did not fit the data as well as a mixture of a
normal prior and a point mass at zero. For the mixture prior, an
empirical Bayes
confidence interval for $\theta_{(i)}$ was constructed and shown
(numerically and asymptotically as $p \rightarrow\infty)$ to have
empirical Bayes confidence coefficient at least $1-\alpha$.

Further, combining $k$ empirical Bayes $1-\alpha/k$ confidence
intervals for $\theta_{(i)}$, $i\in S$, where $S$ consists of~$k$
indices of the selected $\theta_{(i)}$'s, yields a simultaneous
confidence set (rectangle) that has empirical Bayes coverage
probability above the nominal $1-\alpha$ level. Furthermore, their
sizes could be much smaller than even the naive rectangles (which
ignore selection and hence have poor coverage). This can also lead to a~more powerful test.

%with coefficient $(1-\alpha)^k$. The $k$
%high-dimensional rectangle has much smaller length than the
%corresponding rectangle $X_{(i)}\pm c_p\sigma$ where $c_p$ is such
%that $P(\max_{1\le i\le p}|Z_i|<c_p)=1-\alpha$. (Such choice of
%$c_p$ obviously guarantee that $\theta_{(i)}\pm c_p\sigma$ would
%have valid frequentist and Empirical Bayes confidence coefficient.
%Amazingly, Qiu and Hwnag's interval for $\theta_{(i)}$ is also
%shorter than the naive interval $X_{(i)}\pm c_1\sigma$, where $C_1$
%is defined as $c_p$ with $p=1$. Anologous statement similar to the
%last two sentences hold when $\sigma^2$ is replaced by an estimator
%that is proportional to a chi-squared random variable times
%$\sigma^2$.

%The simultaneous $k$-dimensional can be applied to testing
%hypothesis involving random parameter $\theta_{(i)}$ and is shown to
%be more powerful than that corresponding to the valid simultaneous
%set using $q_p$. See Section 4 of Qiu (2004).

%s5.3 ###
\vspace*{-1pt}\subsection{Shrinking Means and Variances}\vspace*{-1pt}

Thus far, we have only discussed procedures that shrink the sample
means, however, confidence sets can also be improved by shrinking
variances. In Section \ref{sec:varshrink} we saw how to construct
improved intervals for the variance. In Berry (\citeyear{Berry94}) it was shown that
using an improved variance estimator can slightly improve the risk of
the Stein point estimator (but not the positive-part). Now we will see
that we can substantially improve intervals for the mean by using
improved variance estimates, when there are\break a~large number of variances
involved.

%Thus far, all the papers in the literature regarding estimation or
%confidence set construction assumed that the variances $\sigma_i^2$
%are either known or estimated by an sample variance $S_i^2$ based on
%the $i$-th population. It would seem likelihood that one can further
%improve this type of procedure by shrinkage the sample variance.
%Testing procedures based on shrinking $S_i^2$ have been proposed for
%microarray. See Thusher et al. (2001 known as SAM technique),
%L$\ddot{o}$nnstedt and Speed (2002), Wright and Simon (2003), Smyth
%(2004), Cui et al. (2005), Lo and Gottardo (2007) and Hwang and Liu
%(2007). In particular, the last paper uses a testing statistic that
%shrinks both the sample means and sample variances.

Hwang, Qiu and Zhao (\citeyear{Hwang08}) constructed empirical Bayes confidence
intervals for $\theta_i$ where the center and the length of the interval
are found by shrinking both the sample means and sample variances.
They took an approach similar to He (\citeyear{He92}), except that the task is
complicated by putting yet another prior on $\sigma_i^2$. The prior
assumption is that $\log{\sigma_i^2}$ is distributed according to a
normal distribution (or~$\sigma_i^2$ has an inverted gamma distribution).
In both cases, their proposed double shrinkage confidence interval maintains
empirical Bayes coverage probabilities above the nominal level, while
the expected length are always smaller than the $t$-interval or the
interval that only shrinks means. Simulations show that the
improvements could be up to 50\%.\vadjust{\goodbreak}

%The distribution of $\log{S_i^2/\sigma_i^2}$ is
%also assumed to be normal. The normal model combining with two
%lognormal distributions is called the lognormal model. The Empirical
%Bayes confidence interval $I_i^{SS}$ constructed is shown to be
%better than $I_i^{SS_g}$, which is constructed using the more
%traditional assumption that $S_i^2/\sigma_i^2\sim\chi_d^2/d$ and
%$\sigma_i^2\sim inverse gamma$, called the inverse gamma model. Here
%5the superscript "SS" stands for shrinking both the sample means and
%variances.

%In the above construction, difficulty of truncation in radius is
%encountered which is handled similar to Qiu and Hwang (2007). Also
%it was discovered that the center (or the point estimator) of the
%interval need to be truncated so it never becomes zero. See the
%similar discussion after (\ref{sec:interval:He}) on Morris'
%interval.

The confidence intervals constructed %using the lognormal model and
%the inverse gamma model all both
are shown to have empirical Bayes
confidence coefficient close to $1-\alpha$.
%However, simulations
%also show that $I^{SS}$ has smaller average length than $I^{SS_g}$,
%although in applying to a set of data (Choe et al. 2005) where the
%"true means" are known, they both perform well in the "coverage
%probability" and "average length".
In all the numerical studies,
including extensive simulation and the application to the
data sets, the double shrinkage procedure %$I^{SS}$
performed better than the single shrinkage intervals (intervals that
shrink only one of the sample means or sample variances but not
both) and the standard $t$ interval (where there is no shrinkage).

%s6 ###
\section{Discussion}

The confidence sets that we have discussed broadly fall into two
categories: those that are explicitly defined by a center and a radius
(such as Berger, \citeyear{Berger80}, or Casella and Hwang, \citeyear{Casella83}), and those that are
implicit (such as Tseng and Brown, \citeyear{Tseng97}). For experimenters, the
explicitly defined intervals may be slightly preferred.

The improved confidence sets typically work because they are able to
reduce the volume of the $x$-section (the confidence set) without
reducing the volume of the $\theta$-section (the acceptance region). As
the coverage probability results from the $\theta$-section, the result
is an improved set in terms of volume and coverage.

Another point to note is that most of the sets presented are based on
shrinking toward zero. Moreover, the improved sets will typically have
greatest coverage improvement near zero, that is, near the point to
which they are shrinking. The point zero is, of course, only a
convenience, as we can shrink toward any point $\mu_0$ by translating
the problem to $x-\mu_0$ and $\theta-\mu_0$, and then obtain the
greatest confidence improvement when $x-\mu_0$ is small. Moreover, we
can shrink toward any linear subset of the parameter space, for
example, the space where the coordinates are all equal, by translating
to $x-\bar x {\bf1}$ and $\theta-\bar\theta{\bf1}$, where $
\mathbf{1}$ is a vector of $1$s. This is developed in Casella and
Hwang (\citeyear{Casella87}).

The Stein effect, which was discovered in point estimation, has had
far-reaching influence in confidence set estimation. It has shown us
that by taking into account the structure of a problem, possibly
through an empirical Bayes model, improved point and set estimators can
be constructed.

\begin{appendix}

%s8 ###
\section{Proof of Dominance of~$C^+$}\label{app:hc}

Hwang and Casella (\citeyear{Hwang82}) show that $P_\theta
(\theta\in C^+)$ is decreasing in $|\theta|$, and hence has minimum
$1-\alpha$ at $|\theta|=\infty$. The proof is somewhat complex, and
only holds for $p \ge4$. Hwang and Casella (\citeyear{Hwang84})\vadjust{\goodbreak} found a simpler
approach, which extended the result to $p=3$. We outline that approach
here.

For the set $C^+ = \{\theta\dvtx | \theta- \delta^+
(\mathbf{x})|
\le c\}$, the following lemma shows that we do not have to worry
about $|\theta|<c$.
\begin{lemma}
For $X \sim N(\theta, I)$ and every $a>0$ and $|\theta|<c$,
\[
P_\theta\bigl(\theta\dvtx | \theta- \delta^+ (X)| \le c\bigr)
\ge P_\theta
(\theta\dvtx | \theta- X| \le c).
\]
\end{lemma}

\begin{pf}
The assumption $|\theta|<c$ implies that $0 \in C_\theta^0$, the
$\theta
$-section (acceptance region). Therefore, by the convexity of $C_\theta^0$,
\[
x \in C_\theta^0 \quad \Longrightarrow\quad \delta^+(x) \in C_\theta^0
\]
since $\delta^+(x)$ is a convex combination of $0$ and $x$. Finally,
since $\delta^+(x) \in C_\theta^0$, we then have $|\delta
^+(x)-\theta|
\le c$ so $C_\theta^0 \subset C_\theta^+$ and the theorem is proved.
\end{pf}

It is interesting that, even though the confidence sets (the $x$-sections) have
exactly the same volume; for small $\theta$ the
$\theta$-section of the $\delta^+$ procedure contains the $\theta$-section of the usual procedure.

In addition to not needing to worry about $|\theta| <c$, there is a
further simplification if $|\theta| \ge c$. If $|\theta| \ge c$, the
inequality $|\theta- \delta^+(x)| \le c$ is equivalent to
\[
|\theta- \delta^+(x)| \le c\quad  \mbox{and}\quad  |x|^2 \ge a,
\]
which allows us to drop the ``$+$.'' Note that if $|\theta| >c$ and $|x|^2
< a$, then $|\theta- \delta^+(x)| > c$.

Last, we note that if $a=0$, then the two procedures are exactly the
same and, thus, a sufficient condition for domination of $C_x^0$ by
$C_\delta^0$ is to show that
%
%e17 ###
%
\begin{equation}\label{eq:HCsufficient}
\frac{d}{da} P_\theta(\theta\in C_\delta^+) >0
\end{equation}
for every $|\theta|>c$ and $a$ in an interval including $0$.
The inequality (\ref{eq:HCsufficient}) was established in Hwang and
Casella (\citeyear{Hwang84}) through the use of the polar transformation $(x, \theta)
\rightarrow(r,\beta)$, where $r=|x|$ and $x^\prime\theta=
|x||\theta
|\cos(\beta)$, so $\beta$ is the angle between $x$ and $\theta$. The
polar representation of the coverage probability is differentiable in
$a$, and the following theorem was established.
\begin{theorem}
For $p \ge3$, the coverage probability of $C_\delta^+$ is higher than
that of $C_x^0$ for every $\theta$ provided $0 < a\le a^\ast$, where
$a^\ast$ is the unique solution to
\[
\biggl(\frac{c^2+(c^2+a^\ast)^{1/2}}{a^\ast}\biggr)^{p-2}e^{-c\sqrt
{a^\ast}}=1.
\]
\end{theorem}

Solutions to this equation are easily computed, and it turns out that
$a^\ast\approx0.8(p-2)$, which does not quite get to the value $p-2$,
the optimal value for $\delta^{\mathrm{JS}}$ and the popular choice for $\delta
^+$. However, the coverage probabilities are very close. Moreover, the
theorem provides a sufficient condition, and it is no doubt the case
that $a=p-2$ achieves dominance.

%s9 ###
\section{The Strawderman Prior}\label{app:straw}
The first proper Bayes minimax point estimators were found by
Strawderman (\citeyear{Strawderman71}) using a hierarchical prior of the form
\begin{eqnarray*}
X|\theta&\sim& {N}_p(\theta, I),\\
\theta|\lambda&\sim& {N}_p\biggl(0, \frac{1-\lambda}{\lambda}I\biggr),\\
\lambda&\sim& (1-a)\lambda^{-a},\quad  0<\lambda\le1,\ 0\le a
< 1.
\end{eqnarray*}

The Bayes estimator for this model is $\mathrm{E}(\theta|x) = [1-\mathrm{E}(\lambda|x)]x$. The function $\mathrm{E}(\lambda|x)$ is a bounded
increasing function of $|x|$, and Strawderman was able to show, using
an extension of Baranchik's (\citeyear{Baranchik70}) result, that for $p \ge5$ the Bayes
estimator is minimax. An interesting point about this hierarchy is that
the unconditional prior on $\theta$ is approximately $1/|\theta
|^{p+2-2a}$, giving it $t$-like tails. These are the types of priors that lead to Bayesian posterior
credible sets with good coverage probabilities.

Faith (\citeyear{Faith78}) used a similar hierarchical model with $\theta\sim
N(0,t^2I)$ and $t^2 \sim\mbox{Inverted } \operatorname{Gamma}(a,b)$, leading to an
unconditional prior on $\theta$ of the form $\pi(\theta) \approx
(2b+|\theta|^2)^{-(p/2+a)}$, the multivariate $t$ distribution. In his
unpublished Ph.D. thesis, Faith gave strong evidence that the Bayesian
posterior credible sets had good coverage properties.

Berger (\citeyear{Berger80}) used a generalization of Strawderman's prior, which is
more tractable than the $t$ prior of Faith, to allow for input on the
covariance structure.
\end{appendix}

\section*{Acknowledgments} Thanks to the Executive Editor,
Editor and Referee for their careful reading and thoughtful
suggestions, which improved the presentation of the material.
Supported by National Science Foundation Grants
DMS-0631632 and SES-0631588.

\end{document}